\documentclass{jps-cp}
\usepackage{txfonts}
\usepackage[crop=pdfcrop,process=auto,cleanup={.dvi,.ps,.pdf,.log}]{pstool}
\title{Electromagnetic Response of Superconducting RF Cavities}
\author{Hikaru \textsc{Ueki},$^*$ Mehdi \textsc{Zarea},$^\dagger$ and J. A. \textsc{Sauls}$^\ddagger$}
\inst{Center for Applied Physics and Superconducting Technologies, 
Department of Physics and Astronomy, 
Northwestern University, Evanston, IL 60208, USA}
\email{$^*$hikaru.ueki@northwestern.edu, 
$^\dagger$zarea.mehdi@gmail.com, 
$^\ddagger$sauls@northwestern.edu}

\recdate{July 31, 2022}

\abst{
Recently reported anomalies in frequency shift of order $\delta f\sim 0.1 -10\,\mbox{kHz}$ for Niobium SRF cavities in a narrow temperature region near $T_c\simeq 9\,\mbox{K}$ are sensitive to the surface preparation and Nitrogen doping.
We developed methods for calculating the surface current response of Nb SRF cavities, as well as the resonant frequency shift and quality factor, as functions of temperature, frequency and disorder based on the Keldysh formulation of the theory for nonequilibrium superconductivity coupled with Maxwell's theory and boundary conditions for the response functions and electromagnetic field.
We show that the anomaly is sensitive to impurity disorder. Our results for the anomaly in the frequency shift are in good agreement with experimental results for Nb with an anisotropic gap on the Fermi surface and inhomogeneous non-magnetic disorder.
We also show that the quality factor as a function of the impurity scattering rate is maximum for in SRF cavities with intermediate disorder, $\hbar/\tau\Delta\sim{\cal O}(1)$ with the maximum $Q$ decreasing with increasing frequency. 
}

\kword{superconductivity, superconducting radio-frequency cavities, quasiclassical theory, Maxwell's equations}

\begin{document}
\maketitle

\section{Introduction}
Niobium superconducting radio-frequency (SRF) cavities have been greatly improved in terms of the quality factor $Q$ by techniques such as Nitrogen doping\cite{gra13,rom14}, however the mechanism behind the increase in $Q$ is not fully understood. 
These high-$Q$ cavities provide new technology platforms for quantum computing~\cite{rom20} and detectors for dark matter (DM) candidates such as axions~\cite{bog19,ber20,gao21}, as well as precision tests of low-energy quantum electrodynamics (QED)~\cite{bog19}.
The sensitivity to the coupling of the axion field to electromagnetism
and the photon-photon interaction predicted by Euler and Heisenberg
depends on the $Q$ of the cavity~\cite{bog19}.
The quality factor of SRF cavities at low temperatures and ultra-low microwave intensities is also of interest for applications in quantum information technology such as quantum memory~\cite{rom20}.
Thus, there is considerable motivation from researchers spanning accelerator physics to quantum computing and sensing to develop a deeper understanding of the mechanisms limiting $Q$ and the lifetime of microwave photons in SRF cavities.

Recent insight comes from precision measurements of the temperature dependence of the frequency of N-doped Nb SRF cavities.
Bafia {\it et al.} report a dip in the frequency of order kHz confined to a very narrow temperature range near the transition temperature \cite{baf21}. 
Anomalies in the frequency shift of Nb coupled to resonators just below $T_c$ were reported earlier but have not been investigated systematically, experimentally or theoretically~\cite{var74,kle94}.
Bafia {\it et al.} provide precision measurements for the temperature dependence of the frequency shift over the full temperature range, including the narrow region of the anomaly just below $T_c$ for cavities with several different resonance frequencies and surface and doping preparations. They report that the minimum value of the frequency shift correlates with the level of disorder based on estimates of the conduction electron mean free path. 

\section{Formalism}

Here we report theoretical results for the {\it a.c.} conductivity, surface impedance, frequency shift and qualify factor of disordered Nb SRF cavities from the Keldysh formulation of the non-equilibrium quasiclassical theory of superconductivity~\cite{rai95} coupled to Maxwell's equations with appropriate boundary conditions for the current response functions and electromagnetic field. 
We compare our theoretical results with the experimental results of Ref.~\cite{baf21} for an N-doped Nb SRF cavity. 
We also show that $Q$ as a function of impurity scattering rate is non-monotonic with a maximum corresponding to the regime of intermediate disorder with $\hbar/\tau\Delta\sim{\cal O}(1)$.    

\subsection{Surface impedance}

The surface impedance of the vacuum-superconducting interface, obtained by solving Maxwell's equations~\cite{Jackson,gur17}, is related to the complex {\it a.c.} conductivity in the local limit,
\begin{align} 
Z_s = Z_0\sqrt{\frac{f}{2i\sigma}}, \label{eq-Zs2}
\end{align}
where $Z_0\equiv 4\pi/c=376.7\,{\rm \Omega}$ is the vacuum impedance, $f=\omega/2\pi$ is the resonance frequency, and $\sigma(\omega)=\sigma_1+i\sigma_2$ is the conductivity.
In the limit $\sigma_{n1} \gg \sigma_{n2}$ for the normal-state, with $\sigma_n=\sigma_{n1} + i \sigma_{n2}$, we obtain the following relations between the surface impedance and complex conductivity
\begin{subequations} 
\begin{align} 
&\frac{R_s}{R_n} 
= \frac{\sigma_{n1}^{1/2}}{(\sigma_1^2+\sigma_2^2)^{1/4}} 
\left[ \cos \left( \frac{1}{2} \arctan \frac{\sigma_2}{\sigma_1} \right) 
- \sin \left( \frac{1}{2} \arctan \frac{\sigma_2}{\sigma_1}  \right) \right]
\,,
\label{eq-Rs}
\\
&\frac{X_s}{X_n} 
= \frac{\sigma_{n1}^{1/2}}{(\sigma_1^2+\sigma_2^2)^{1/4}} 
\left[ \cos \left( \frac{1}{2} \arctan \frac{\sigma_2}{\sigma_1} \right) 
+ \sin \left( \frac{1}{2} \arctan \frac{\sigma_2}{\sigma_1}  \right) \right]
\,,
\label{eq-Xs}
\end{align}
\label{eq-Zs}
\end{subequations} 
where the normal-state resistance ($R_n$) and reactance ($X_n$) are  
\begin{align} 
R_n=X_n=\frac{Z_0}{\omega_p}\sqrt{\frac{\pi f}{\tau}}
\,,
\label{eq-Rn}
\end{align}
and $\omega_p$ is the plasma frequency and $\tau$ is the mean time between scattering events for conduction electrons in the normal state, i.e. the ``relaxation time''.
We use Eq.~(\ref{eq-Rn}) for the normal-state resistance and $\omega_p=9.08\times10^6 \ {\rm GHz}$ \cite{kle94} for Niobium to fix the scale for frequency shift and quality factor. 

\subsection{Frequency shift and quality factor}

Slater considered the electromagnetic (EM) fields in a hollow cavity and derived relations between the frequency shift (quality factor) and surface reactance (resistance) by expanding the EM fields in the basis of eigenfunctions (modes) of an ideal cavity defined as a perfect conductor with infinite {\it d.c.} conductivity~\cite{sla46}.
For SRF cavities in which the interior is vacuum, and there is no insulating layer coating the superconducting surface, the frequency shift and quality factor obtained from Slater's method are given by 
\begin{align} 
\delta f=\frac{f}{2G}(X_n-X_s)\,,\quad Q=\frac{G}{R_s}
\,,
\label{eq-Delta_f_and_Q}
\end{align}
where 
\begin{align} 
G\equiv Z_0\frac{2\pi f}{c}\int_V{\bf H}^2dv\Bigg/\int_S{\bf H}^2da
\,,  
\end{align}
is a geometric factor defined by the magnetic field distribution of the ideal cavity mode $\bf H(\bf r)$, specifically the ratio of the magnetic field energy density integrated over the volume ($V$) inside the cavity and the inner surface ($S$) of the cavity.

\subsection{Conductivity for a disordered isotropic superconductor}

We consider a conventional ``$s$-wave'' superconductor in the weak coupling limit with homogeneous disorder defined by the normal-state transport scattering rate, $1/\tau$. We initially neglect the anisotropy of the gap function defined on the Fermi surface. Then in the long-wavelength limit of Eq.~(68) in Ref. \cite{rai95} the {\it a.c.} conductivity in the linear response limit obtained from the Keldysh formulation of the quasiclassical equations of superconductivity reduces to
\begin{align} 
\sigma (\omega)& = \frac{\sigma_{\rm D}}{i\omega\tau} 
\int_{-\infty}^\infty\frac{d\varepsilon}{4\pi i} \notag \\
\times \bigg\{
&2\tanh\left(\frac{\varepsilon-\omega/2}{2T}\right)
\frac{-\pi}{D^{\rm R}(\varepsilon+\omega/2)+D^{\rm R}(\varepsilon-\omega/2)+1/\tau}
\left[\frac{\varepsilon^2-\omega^2/4+\Delta^2}
{D^{\rm R}(\varepsilon+\omega/2)D^{\rm R}(\varepsilon-\omega/2)}
+1\right] \notag \\
+\bigg[ &\tanh\left(\frac{\varepsilon+\omega/2}{2T}\right)
-\tanh\left(\frac{\varepsilon-\omega/2}{2T}\right)\bigg] \notag \\
& \ \ \ \ \ \ \ \ \ \ \ \ \ \ \ \ \ \ \ 
\times\frac{-\pi}{D^{\rm R}(\varepsilon+\omega/2)+D^{\rm A}(\varepsilon-\omega/2)+1/\tau}
\left[\frac{\varepsilon^2-\omega^2/4+\Delta^2}
{D^{\rm R}(\varepsilon+\omega/2)D^{\rm A}(\varepsilon-\omega/2)}
+1\right] 
\bigg\}
\,,
\label{eq-sigma}
\end{align}
where $\sigma_{\rm D}$ is the {\it d.c.} Drude conductivity, $\omega =2\pi f$ is the resonance frequency, and the retarded (R) and advanced (A) functions $D^{\rm R,A}(\varepsilon)$ are defined by $D^{\rm R,A}(\varepsilon)\equiv\sqrt{\Delta^2-(\varepsilon\pm i\eta)^2}$ with $\eta\to0^+$. N.B. the upper and lower signs correspond to the retarded and advanced functions, respectively. 

\subsection{Inhomogeneous Disorder}

Bafia {\it et al.} measured the frequency shift of Nb SRF cavities very near $T_c$ in detail \cite{baf21}. These macroscopic superconductors are inhomogeneous materials, particularly in terms of the distribution of disorder. For an anisotropic superconductor such as Nb even non-magnetic impurity disorder leads to a suppression of the transition relative to the maximum $T_c$ for pure, single-crystal Nb~\cite{zar22}. Indeed Bafia reports a spread in $T_c$ depending on the location of the measurement of the onset of superconductivity~\cite{bafpc}.
The analysis presented here shows that inhomogeneity in disorder, i.e. in $1/\tau$, and thus the distribution for $T_c$, is important for a quantitative understanding on the anomaly in the frequency shift very near $T_c$. 
In particular, while theoretical results based on homogeneous disorder predict a negative frequency shift of the right magnitude the predicted width of the anomaly in temperature does not accurately describe the experimental width. 
We assume a Gaussian distribution for spread in $T_c$,
\begin{align} 
\rho(T_c)
=\frac{1}{\sqrt{2\pi\mu_2}}{\rm e}^{-\frac{(T_c-\mu)^2}{2\mu_2}}
\,,
\label{eq-GaussianTc}
\end{align}
where $\mu=T_c^{\rm ave}$ and $\mu_2$ are the average and variance, respectively. 
The average of the gap function, 
\begin{align} 
\Delta(T)\equiv\Delta_0\sqrt{\int_{-\infty}^\infty dT_c\,\rho(T_c)\,\tilde{\Delta}^2(T,T_c)}
\,, 
\quad
\tilde{\Delta}(T,T_c)=\tanh\left(\frac{\pi T_c}{\Delta_0}
\sqrt{\frac{2}{3}\frac{\Delta C}{C_n(T_c)}\frac{T_c-T}{T}}\right)
\Theta(T_c-T)
\,,
\label{eq-gap_spread_in_Tc}
\end{align}
determines the temperature dependence of the conductivity at the onset of the superconducting transition. Note that $\Delta_0=\pi e^{-\gamma}T_c^{\rm ave}$ is the gap energy at $T=0$, where $\gamma=0.57721\cdots$ is the Euler--Mascheroni constant, $T_c^{\rm ave}$ is the average of $T_c$, $\Delta C \equiv C(T_c)-C_n(T_c)$ is the jump in the heat capacity at $T_c$ given by $\Delta C=12\,C_n(T_c)/7\zeta(3)$ in the weak coupling limit, $C(T_c)$ and $C_n(T_c)$ denote the superconducting and normal heat capacities at $T_c$, respectively, and $\zeta(3)=1.20205\cdots$ is the Riemann zeta function.

The distribution of disorder is directly related to Eq.~\eqref{eq-GaussianTc} with the connection provided by equation of the suppresion of $T_c$ as a function of the scattering rate $1/\tau$, 
\begin{align} 
\tilde{\rho}(1/\tau)=-\frac{dT_c(1/\tau)}{d(1/\tau)}\frac{1}{\sqrt{2\pi\mu_2}}
{\rm e}^{-\frac{[T_c(1/\tau)-\mu]^2}{2\mu_2}}
\,,
\label{eq-probability_density_tau}
\end{align}
where $\mu$ and $\mu_2$ are the same parameters as those in Eq.~(\ref{eq-GaussianTc}). The functions $T_c(1/\tau)$ and $dT_c(1/\tau)/d(1/\tau)$ are calculated by solving equation for the suppression of 
$T_c$ as a function of disorder and gap anisotropy~\cite{zar22},
\begin{align} 
\ln \frac{T_{c_0}}{T_c}={\cal A}\sum_{n=0}^\infty
\left(\frac{1}{n+\frac{1}{2}}-\frac{1}{n+\frac{1}{2}+\frac{1}{2}\frac{1/\tau}{2\pi T_c}}\right), \label{eq-Tc_eq}
\end{align}
where ${\cal A}\equiv\lim_{T\to T_c}(\langle|\Delta({\bf p})|^2\rangle-|\langle\Delta({\bf p})\rangle|^2)/\langle|\Delta({\bf p})|^2\rangle$ is a dimensionless measure of the gap anisotropy defined with the the Fermi surface average $\langle\cdots\rangle$ normalized as $\langle1\rangle=1$ and ranging from ${\cal A}=0$ (isotropic $s$-wave superconductors) to ${\cal A}=1$ (unconventional superconductors). Note that $T_{c_0}$ is the transition temperature in the clean limit. We use ${\cal A}=0.037$ and $T_{c_0}=9.33\,{\rm K}$ for pure Niobium obtained from LDA and Eliashberg theory~\cite{zar22},
and $dT_c(1/\tau)/d(1/\tau)$ calculated from Eq.~\eqref{eq-Tc_eq}. 
%

\begin{figure}[t]
\begin{center}
\includegraphics[width=\linewidth]{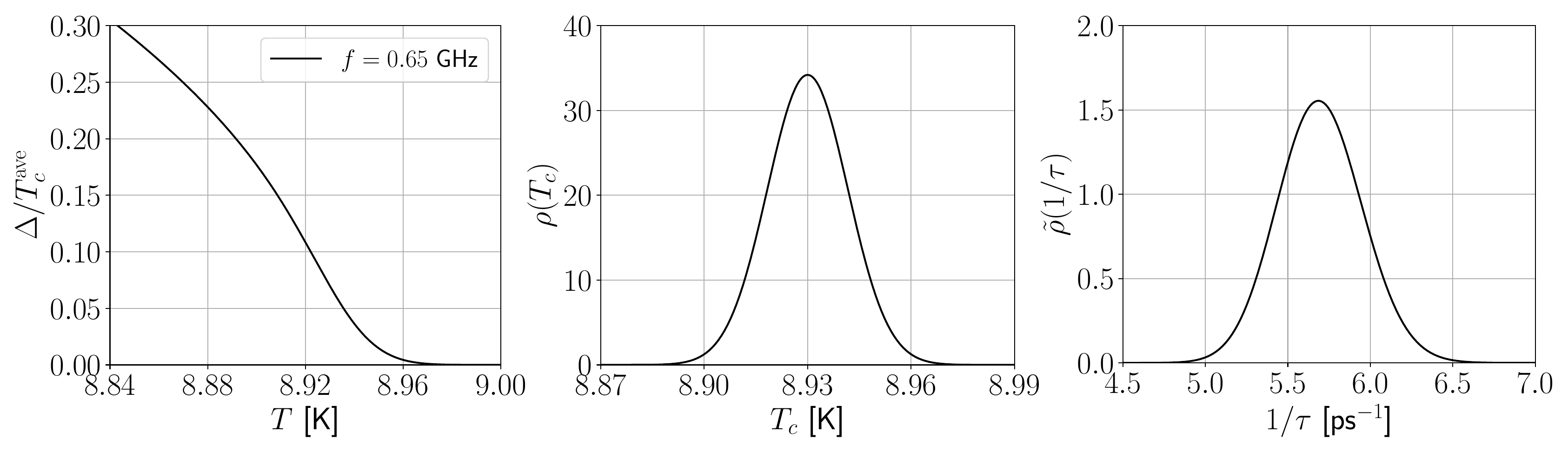}
\caption{
Left: the average energy gap $\Delta(T)$ near $T_c$.
Center: probability distribution for the transition temperature, $\rho(T_c)$,
Right:  probability distribution for the scattering rate, $\tilde{\rho}(1/\tau)$, 
for the N-doped Nb SRF cavity with $f=0.65\,{\rm GHz}$ reported in Ref.~\cite{baf21}.}
\label{fig-inhomogeneity}
\end{center}
\end{figure}

\begin{figure}[t]
\begin{center}
\includegraphics[width=0.6\linewidth]{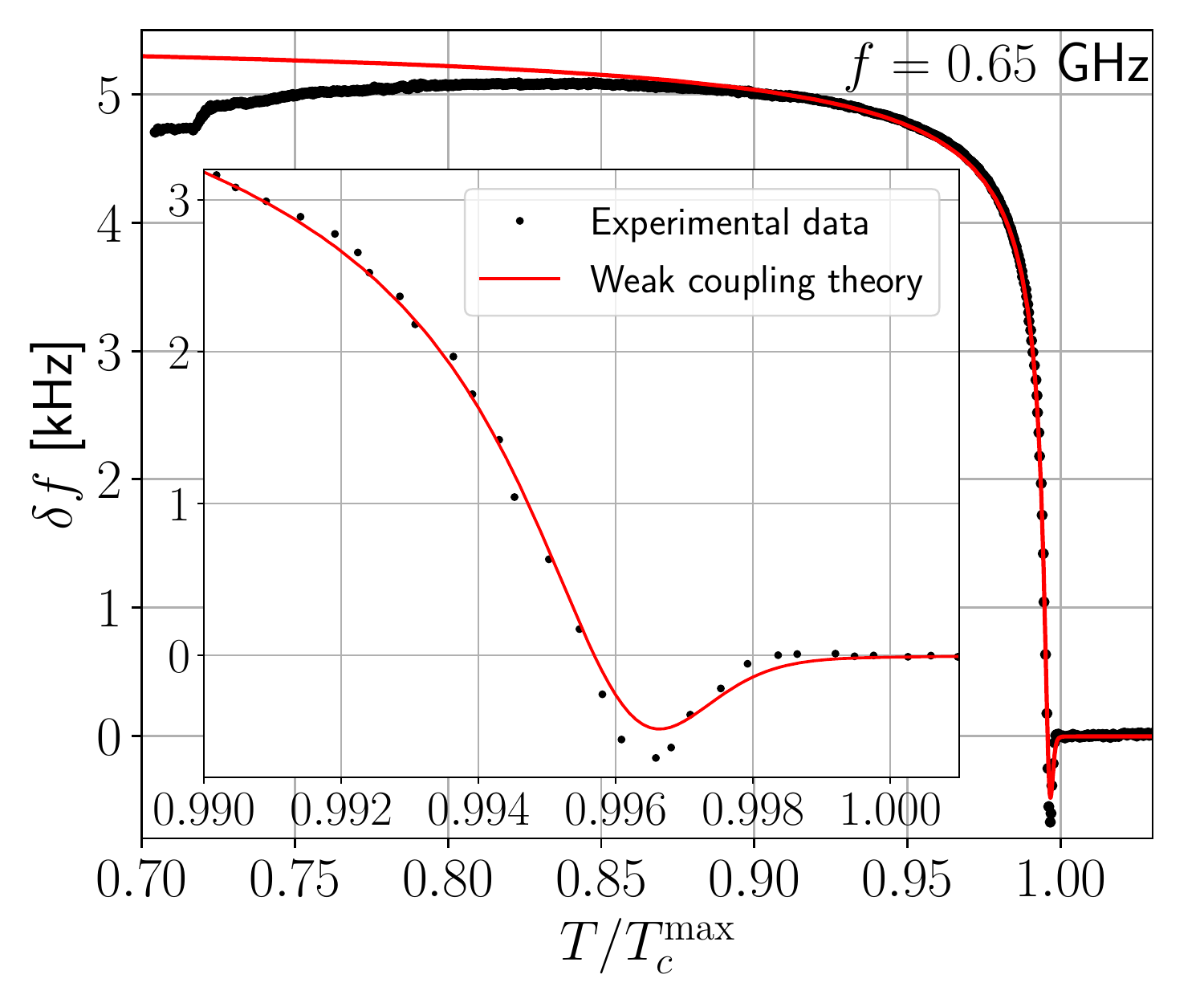}
\caption{Temperature dependence of the frequency shift $\delta f$ 
of the N-doped Nb SRF cavity with $f=0.65\,{\rm GHz}$. 
The black dots are values reported by Bafia {\it et al.}, 
and the red lines is the theoretically calculated shift as described in the text.
The inset highlights the anomaly in the region very near $T_c$.}
\label{fig-Bafia}
\end{center}
\end{figure}

\section{Results}

Figure~\ref{fig-inhomogeneity} shows the calculated temperature dependence of the gap near $T_c$ obtained from the distribution of $T_c$, and the corresponding probability distribution for the scattering rate, $1/\tau$, for the N-doped Nb SRF cavity with $f=0.65\,{\rm GHz}$~\cite{baf21}.
We have introduced the parameters $T_c^{\rm max}$ and $T_c^{\rm min}$ and expressed the average and variance in the Gaussian distribution of $T_c$ as $\mu=T_c^{\rm ave}\equiv(T_c^{\rm max}+T_c^{\rm min})/2$ and $\mu_2=[(T_c^{\rm ave}-T_c^{\text{min}})/3]^2$, respectively.
Note the slow rise in the gap due to the spread in $T_c$, which affects the width of the anomaly in the frequency shift. We can evaluate Eqs.~(\ref{eq-sigma}) and (\ref{eq-Rn}) using the average value of the scattering rate given by $\tau^{\rm ave}T_{c_0}=0.214$. Note also that the average of $\tau$ is almost the same as the inverse of the average of $1/\tau$. Thus, we can calculate the average of both $\tau$ and $1/\tau$ by solving Eq. (\ref{eq-Tc_eq}) at $T_c=T_c^{\rm ave}$.
These results are inputs to our calculation of the temperature dependence of the frequency shift of the same cavity obtained from Eqs.~(\ref{eq-Delta_f_and_Q},\ref{eq-Xs},\ref{eq-Rn},\ref{eq-sigma}) with the geometric factor $G=255\,{\rm \Omega}$~\cite{bafpc}. 
The parameters, $T_c^{\rm max}=8.965\,{\rm K}$ and $T_c^{\rm min}=8.895\,{\rm K}$, were obtained by optimizing the agreement of the calculated frequency shift with the experimental data.
The comparison between our theoretical result and the experimental data reported by Bafia {\it et al.}~\cite{baf21} is shown in Fig.~\ref{fig-Bafia}, both for the full temperature range as well as the narrow temperature range of the negative frequency shift near the onset of superconductivity (inset).
The resulting distribution for the transport scattering rate corresponds to inhomgeneous disorder centered around an average scattering rate correspoding to intermediate disorder, $\hbar/\tau_{\rm ave}\Delta\sim{\cal O}(1)$. As for the best fit values of the parameters $T_c^{max,min}$ they are in reasonably good aggreement with the experimental measures of the onset of superconductivity reported by Bafia {\it et al.}, $T_c^{\rm max}=9.005\,{\rm K}$ and $T_c^{\rm min}=8.975\,{\rm K}$. 
In addition, we obtain a normal-state surface resistance of $R_n=4.471\,{\rm m\Omega}$, compared to the experimentally reported value of $R_n=4.364\,{\rm m\Omega}$.

To understand the origin of the negative frequency shift confined near $T_c$ it is useful to express the conductivity near $T_c$ as a perturbative correction to the normal-state conductivity, $\sigma_1=\sigma_{1n}+\delta\sigma_1$ and $\sigma_2=\sigma_{2n}+\delta\sigma_2$.
In the limit $\sigma_{1n}\gg\sigma_{2n}$, relevant to N-doped Nb SRF cavities, we obtain a frequency shift very near $T_c$ given by $\delta f=(fR_n/4G\sigma_{1n})(\delta\sigma_1-\delta\sigma_2)$. For strong to intermediate disorder the leading corrections to the conductivity satisfy $\delta\sigma_1-\delta\sigma_2<0$ very near $T_c$ leading to a negative frequency shift. As the temperature drops the gap opens, the dissipative normal fraction is suppressed and the real part of the conductivity increases leading a cross-over near $T_c$ to a positive frequency shift.  
At higher frequencies, e.g. $f\approx 60\,{\rm GHz}$, the onset of the negative shift at $T_c$ is more pronounced due to an increased window for pair-breaking by the microwave field, i.e. $\hbar\omega>2\Delta(T)$ for $T_c-T\ll T_c$, even for frequencies $\hbar\omega\ll 2\Delta_0$~\cite{uek22}.

\begin{figure}[t]
\begin{center}
\includegraphics[width=\linewidth]{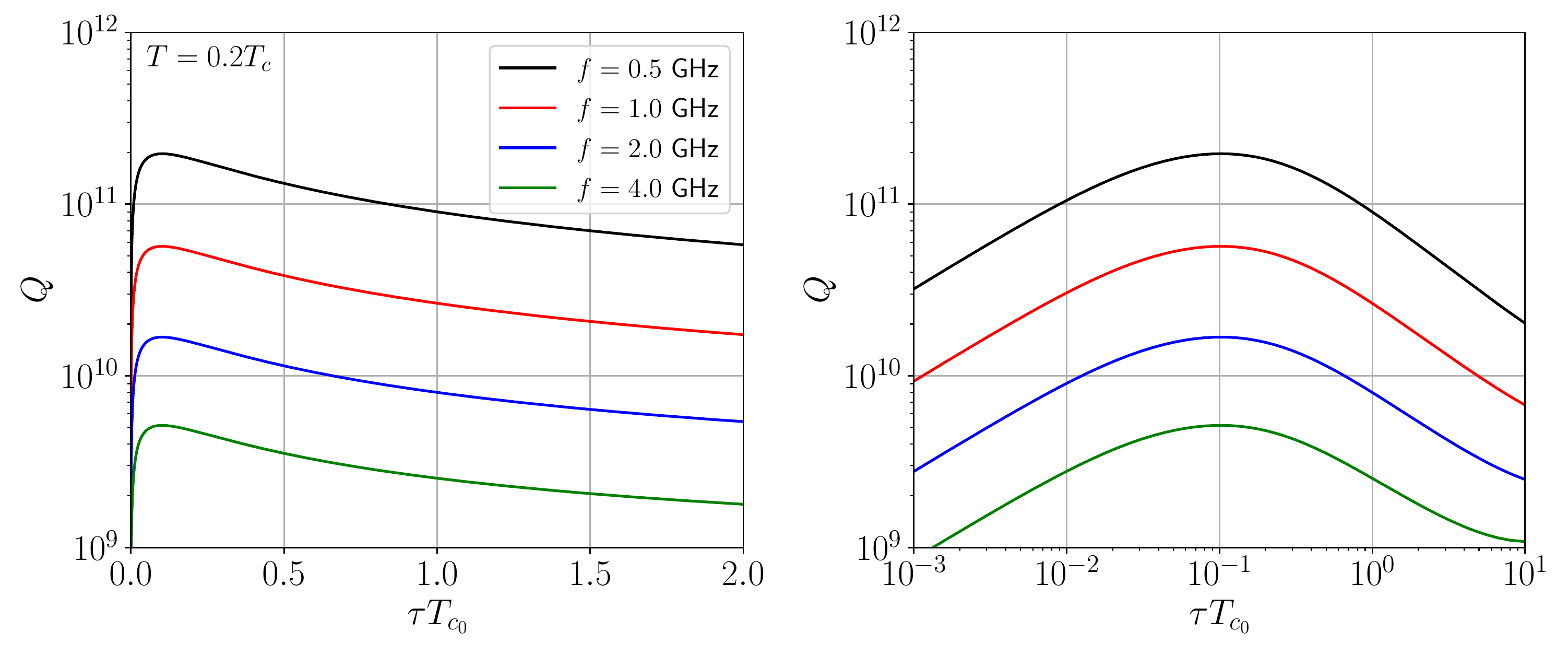}
\caption{Theoretical prediction for $Q$ at $T=0.2T_c$ as a function of $1/\tau$ and cavity frequency.
The left (right) panel is a linear (logarithmic) scale for disorder.}
\label{fig-Delta_f_and_Q}
\end{center}
\end{figure}

The theoretical framework also allows us predict the effect of disorder on $Q$. Figure \ref{fig-Delta_f_and_Q} shows the results for $Q$ as a function of the relaxation time for $T=0.2T_c$, and $G=250\,{\rm \Omega}$. The quality factor is non-monotonic with a maximum corresponding to SRF cavities in the intermediate disorder limit, $\tau T_{c_0}\approx 0.1$.The quality factor is suppressed for stronger disorder due to pair breaking by the interplay between disorder and the screening current. 
In summary we present the first theoretical explanation for the negative frequency shift anomaly observed in disordered Nb-based resonators. Our results also demonstrate the sensitivity of the cavity frequency to the magnitude and spatial inhomogeneity of the impurity potential. An extension of this analysis to SRF cavities with different levels of disorder over a wide range of frequencies is reported in Ref.~\cite{uek22}.

\section*{Acknowledgments}

We thank Daniel Bafia, Anna Grassellino, Alex Romanenko and John Zasadzinski for many discussions. HU and MZ were supported by NSF Grant PHY-1734332. JAS was supported by DOE Office of Science, National Quantum Information Science Research Centers, Superconducting Quantum Materials and Systems Center under contract DE-AC02-07CH11359.

\end{document}